\PassOptionsToPackage{dvipsnames,svgnames}{xcolor}
\documentclass[journal]{vgtc}   

\onlineid{0}

\vgtccategory{Research}

\vgtcpapertype{system}

\PassOptionsToPackage{naturalnames}{hyperref}

\title{\texorpdfstring{Path-based Design Model\\ for Constructing and Exploring Alternative Visualisations}{Path-based Design Model for Constructing and Exploring Alternative Visualisations}}

\author{%
 \authororcid{James Jackson}
    {0000-0002-9383-5888},
 \authororcid{Panagiotis~D. Ritsos}
    {0000-0001-9308-3885}, 
 \authororcid{Peter W.\ S.\ Butcher}{0000-0002-3361-627X} and
    \authororcid{Jonathan~C. Roberts}
    {0000-0001-7718-3181} 
 }	
\authorfooter{
	\item J. Jackson was with the School of Computer Science and  Engineering, Bangor University, UK. Now with \href{https://exadev.io}{ExaDev}.
    \item P.D. Ritsos, P.W.S Butcher and J.C. Roberts, are with the School of Computer Science and  Engineering, Bangor University, UK.
	E-mail:
 \{p.ritsos;p.butcher,j.c.roberts\}@bangor.ac.uk.}

\abstract{%
We present a path-based design model and system for designing and creating visualisations. Our model represents a systematic approach to constructing visual representations of data or concepts following a predefined sequence of steps. The initial step involves outlining the overall appearance of the visualisation by creating a skeleton structure, referred to as a flowpath. Subsequently, we specify objects, visual marks, properties, and appearance, storing them in a gene. Lastly, we map data onto the flowpath, ensuring suitable morphisms. Alternative designs are created by exchanging values in the gene. For example, designs that share similar traits, are created by making small incremental changes to the gene. Our design methodology fosters the generation of diverse creative concepts, space-filling visualisations, and traditional formats like bar charts, circular plots and pie charts. Through our implementation we showcase the model in action. As an example application, we integrate the output visualisations onto a smartwatch and visualisation dashboards. In this article we (1) introduce, define and explain the path model and discuss possibilities for its use, (2) present our implementation, results, and evaluation, and (3) demonstrate and evaluate an application of its use on a mobile watch.
}


\teaser{
\centering
  \includegraphics[alt={Screenshot of the Genii tool in action, with added labels to describe how to control the tool. The main screenshot shows three sections, a menu of visualisation properties (on the left); a panel of created genes (centre), and the final panel of rendered visualisations (right).},width=\columnwidth]{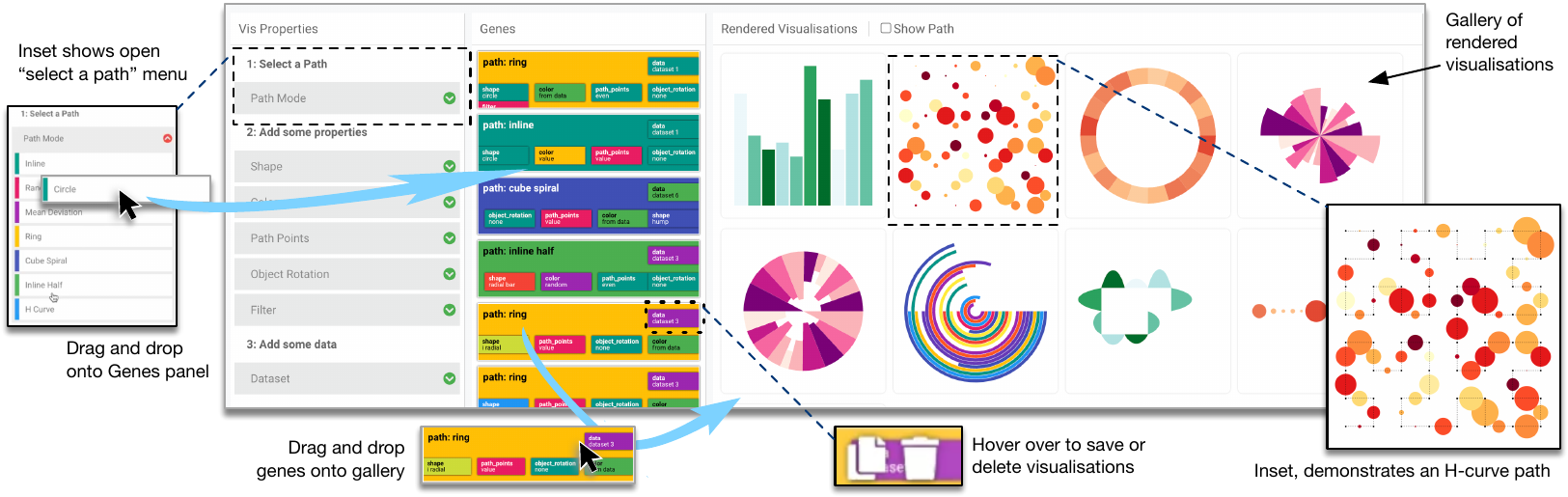}
  \caption{
  The Genii visualisation designer that demonstrates our flowpath model. 
  Individuals define their own path or choose predefined flowpaths (left panel), drag and drop the visualisation properties into the gene panel (middle), which are rendered onto the gallery (right). Users can either create a new gene which adds a new image to the gallery, or edit parameters (through drag and drop) to adapt current visualisations. Crafted visualisations can be exported and used in other applications. 
  }
    \label{fig:teaser}
} 

\graphicspath{{figs/}{figures/}{figures/}{pictures/}{./}} 

\usepackage{mathptmx}                  
\PassOptionsToPackage{warn}{textcomp}  

\usepackage{enumitem}
\usepackage{listings}
\usepackage{tikz}
\usepackage{pgfplots}
\usepackage{pgfplotstable}
\pgfplotsset{compat=1.14}
\usepgfplotslibrary{statistics}
\usepackage{sansmath}

\newlength{\mylen}
\AtBeginDocument{\setlength{\mylen}{\fontcharht\font`X}}
\definecolor{myGray}{RGB}{160,160,160}
\definecolor{editorTeal}{RGB}{32,178,170}
\definecolor{editorBluish}{RGB}{0,191,255}
\definecolor{diagramBlue}{RGB}{144,211,245}
\definecolor{diagramYellow}{RGB}{250,212,93}
\definecolor{diagramOrange}{RGB}{244,155,127}
\definecolor{diagramGreen}{RGB}{208,230,187}
\definecolor{diagramGrey}{RGB}{233,230,229}

\PassOptionsToPackage{font={sf,footnotesize}}{caption}

\begin{document}
\firstsection{Introduction}
\maketitle

\renewcommand*{\backref}[1]{
  %
}

Alternative visualisations are useful but challenging to create. Employed throughout the visualisation domain, they range from sketched planning to explore potential design solutions, to multiple views to explore alternative viewpoints. They offer different perspectives, inspire creativity, validate insights, and help communicate varied narratives. When people set out to create a new visualisation, they often explore numerous design ideas, making many quick sketches on paper. 
For example, by adhering to methods such as the Five Design-Sheets approach, people produce numerous quick sketches to outline their concepts~\cite{RobertsHeadRitsos2016}. They draw a horizontal line to represent an axis, and place three or four rectangles above it to depict a bar chart. Draw two perpendicular lines to represent an X and Y axis, and add a few circles within the space to depict a scatter plot. Or draw a simple rectangle to denote a computer monitor.

\begin{figure}[h!]
    \centering
    \includegraphics[alt={Figure showing sketches of humans that show Reilly line of actions. The line of action is a red line that goes from the head to the foot.},width=\columnwidth]{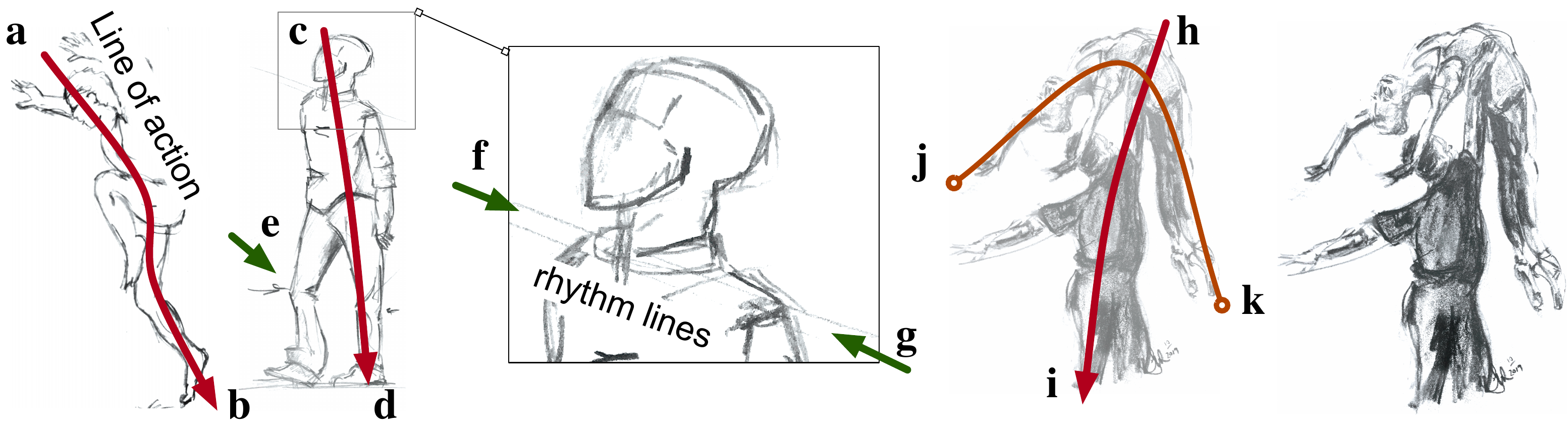}
    \caption{Reilly lines of action~\cite{Matesi2006}. 
    The method is a constructive approach that emphasises primary lines of action, a$\rightarrow$b, c$\rightarrow$d, h$\rightarrow$i, j$\rightarrow$k, and secondary forms (rhythm lines), e$\rightarrow$, and f$\rightarrow$g.}
     \label{FIG:Reilly}
\end{figure}

Individuals are sketching skeletal illustrations to represent various visualisation concepts. They begin by outlining shapes such as rectangles or curves and straight lines to capture the fundamental structure and form of the object being depicted. 
Similar skeletal outlines are employed in other areas of design. This approach reflects a fundamental principle in figure drawing: starting with the essence or skeleton of the subject before adding details. Frank Reilly explained this process as starting to draw the human form from its ``line-of-action''~\cite{Matesi2006,Faragasso2004}. It involves drawing lines that represent the primary movement or flow of the figure's pose, before detailing the rest of the figure, see~\cref{FIG:Reilly}.
The concept of creating a skeleton serves as a powerful strategy for visualisation design. Different paths can be imagined, from x-axis, circles, zigzags, to space-filling curves such as Hilbert, Moore or Peano curves.

\begin{figure}[!h]
    \centering
\includegraphics[alt={Schematic showing five stages of path creation to generate an angled bar chart.},width=\columnwidth]{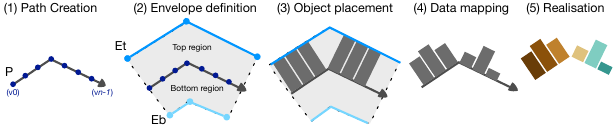}
  \caption{To create the visualisation, first define the sequence, describe the envelope, place objects, map data and add any final filters to realise the visualisation. We represents the points $P$ as a sequence of vertices $v_0$ to $v_{n-1}$ (1). This acts as a skeleton that forms the supporting structure of a visualisation. The envelope (2) acts to control the scope of the placed objects ($E_t$ to $E_b$). Data is mapped along the sequence of points (4), where in this example data is applied to size the objects to create an angled bar chart. Finally colour is mapped to the sequence data to complete the visual design (5).
  }
    \label{fig:flowgraph}
\end{figure}

Our conceptual idea is straightforward yet impactful: establish a skeleton, represented by a flowgraph, to provide the structure for the visualisation. Encode the parameters of the visualisation design using a genetic metaphor, where various components such as symbols, colours, and textures represent genes. Swap in and out different aspects of this `gene', such as symbols, colour, textures, to generate diverse design variations. The flowpath acts as a backbone to the visualisation, on which we constrain the path through an envelope description, which confines the scope of the visuals. We map objects along the path in sequence, and map data values on visual attributes of the objects to encode data. The power with this strategy is to explore designs quickly by exchanging values in the gene. Effects and filters can be mapped along the path, such to merge or intersect overlapping objects, or accumulating transparency of overlapping objects, or morphing them together in some way. \Cref{fig:flowgraph} outlines this concept, exemplified by a simple bar chart on an angle. The flowpath is defined by a contiguous set of vertices $v_0$ to $v_{n-1}$, but does not need to be continuous in space, as it can contain spatial jumps, which allows visual separate designs (such as small-multiples) to be created. In the case of the bar chart, rectangular objects placed between the vertices, which are scaled and coloured by the data. Other visualisation forms are made by changing the sequence order, locations of points, form of objects, envelop, alignment of the actual object (middle, top, bottom etc.), and applied filters on the placed objects. 

While the idea is simple, the challenge is to extend this principle and to develop a framework around this model, that allows users to express their creativity and to be able to easily make their own visualisation designs. 
This work substantially extends our poster presentation at IEEE VIS 2018~\cite{Jackson-et-al-Poster-VIS2018} and workshop paper~\cite{Jackson2019}. 
We make four contributions: 
\textbf{(1) A path-based model} for the creation of 
visualisation designs, visualisations are structured on this backbone skeleton path, \cref{SEC:model}. We present the model such that other people can implement and expand it.
\textbf{(2) The Genii system } implementation contains a component library, renderer and builder application, enabling users to create visualisations by constructing genes through the path-based design model.  Outputs from the builder are displayed in \cref{fig:teaser,fig:caseStudy,FIG:GeniiPict} and \cref{SEC:implementation}. Additionally, we showcase a mobile watch implementation that how the system can create visualisations for a smartwatch~\cref{fig:mobileTool}. 
\textbf{(3)~Several case studies}, where we re-create published  visualisations  using our path-based model~\cref{SEC:case_studies,fig:caseStudy}, and  
\textbf{(4)~an evaluation} of the tool and model, \cref{sec:Evaluation}.

\section{Background}
We have long advocated the use of alternative visualisations for various purposes. In the design phase, it enables individuals to explore various potential solutions~\cite{RobertsHeadRitsos2016}. In exploratory visualisation, different forms can focus on specific tasks or facilitate particular interactions~\cite{Roberts2007}. Meanwhile, in explanatory visualisation, different views can help individuals understand the information more effectively~\cite{RobertsETAL2021ARTEMUS}. 
Several ideas converged to inspire this work, including alternative design, the skeleton pathway, and design similarity.

Our classes on creative visualisation design employ the Five Design-Sheets (FdS) approach~\cite{RobertsHeadRitsos2016}, where multiple sketches are an essential part of the initial sheet. A key principle of this creative design process is encouraging individuals to produce numerous \textbf{alternative} rapidly drawn sketches during the design planning phase. This method allows for a more effective exploration of a wide range of potential solutions.
In our sketching approach, we emphasise the importance of drawing confident yet simple lines. These initial lines serve as the foundational framework, akin to a \textbf{skeleton}, upon which more detailed and refined elements can be built. This method helps designers establish the basic structure and proportions of their work early on, providing a clear guide for subsequent stages of the design process. By focusing on strong, assured strokes, individuals can better visualise the overall composition and make necessary adjustments before committing to more intricate details. 
In another course, we provide general sketching skills with an emphasis on drawing human figures. In art education there are several approaches. Bridgman~\cite{Bridgman1971} emphasises the masses of the body and their connections. Loomis~\cite{Loomis1971} works with proportions and builds up the body from simple shapes --- there are similarities between ideas of sketching by Loomis~\cite{Loomis1971} and scaffold shapes, with the diatoms approach by Brehemer et al.\ \cite{BrehmerETALDiatoms2022}.  We however focus on Frank Reilly's method~\cite{Matesi2006,Faragasso2004}, which guides students to draw simple lines (flow lines of action) along with secondary structures (rhythm lines), see~\cref{FIG:Reilly}. 
Artists are encouraged to first sketch the line-of-action before adding in the body, legs, head and other forms. It is a constructive approach. By first defining the lines of action the human shapes are more expressive, and the sketch is more convincing.  Simple shapes can be defined by single lines, e.g, a line top to bottom when someone is standing \cref{FIG:Reilly}(c$\rightarrow$d) or an S-shape when someone is leaping (a$\rightarrow$b). Complex shapes can be expressed with multiple lines of action as (h$\rightarrow$i) and (j$\rightarrow$k). As we examine these sketches, our attention is instinctively drawn to the principal lines of action. We imagined building visualisations by similar methods.

Our third source of inspiration, \textbf{similarity}, and design modification. This goal  emerged during consultations with a company specialising in nursing applications. They aimed to create alternative visualisations for presenting patient health data, ensuring consistency in visual representations for patients with similar ailments while being distinct enough to differentiate between individual types of ailments. The approach of similarity and design modification offers many benefits.  Modifications allow designers to experiment with variations without losing the core essence of the design, balancing innovation with coherence. This method enables the creation and evolution of a suite of ideas, allowing people to explore and refine concepts effectively.
\color{black}

\vspace{-2mm}\subsection{Design goals}
\label{SEC:DesignGoals}

\newcommand{\codeword}[1]{\small{\texttt{\textbf{#1}}}\normalsize}
We define six design goals that define specific objectives and intentions to guide the creation and implementation of our path-based design model for alternative visualisations.

Our first goal is to develop a method to \textbf{compose} the visualisation from a series of straightforward steps.  This is beneficial because it enables a module design, where individual components can be developed, tested and maintained separately, and it promotes re-usability and scalability.
We explored the use of higher-order functions, like such as \textit{map}, from declarative and functional languages.  \textit{Map} is used to apply a user-defined \textit{function} onto each element of a \textit{list}. Where for instance, \codeword{const l = [1, 2, 3]} makes a list  \codeword{l}, and \codeword{const r = l.map((v, k) => v * k)} multiplies down the list, and the result \codeword{r} becomes a list \codeword{[0, 2, 6]}. We conceptualised that a \textit{path} is a sequence of coordinates, where developers would apply visual design elements onto every facet of the path. Data is mapped to each object along the path. The visual components are combined to form a sophisticated visual design. 

We require mechanisms to enable \textbf{paths} of different shapes, and allow them to be continuous and non-contiguous. This is beneficial to allow flexibility in data presentation, varied visual interpretations, and support different visualisation types. One example use of paths, is from the data-sonification paths of Roberts and Franklin~\cite{RobertsFranklin2005}. They use a path to scan the visualisation, map a kernel each position, which is turned to sound. Another example is the PATH definition in the SVG web format, which can describe non-contiguous paths \codeword{<path d="M0 0 L0 100 L 100 100 Z M 200 0 L 100 100 L 200 100 Z" />}. Consequently, linear paths would make a bar chart, whereas circular paths could be used to create a dial visualisation. 

Secondly, we recognised the importance of having self-contained visualisations that render within a \textbf{viewport}. The visualisation can encapsulate all necessary components (data, styling, interactions) within itself, makes the system portable, can be optimised for an available screen real estate, and offers portability as it can adapt for different viewport sizes. For instance, the nursing application visualisations would need to be sized appropriately to appear at the top of other text descriptions of the patient. The visualisation for our smartwatch (977 $mm^2$ for an Apple Watch 5, 44mm, or a few centimetres in physical dimensions) represents a small viewport. Thus a design goal is to allow a range of paths to be created, and constrain them to a viewport. Our approach is to clip the visualisations using an envelope (see \cref{sec:Envelope}).

Third, we require a \textbf{deterministic} system. A deterministic system ensures that given the same input or conditions, the visualisation will always produce the same output. It means the system is consistent, easier to debug and test because the behaviour is predictable, and results can be compared. This is crucial for scientific research, where reproducible of results is essential for validation and verification. While random paths could yield interesting and unique designs, they would vary potentially with each run. Our solution is to use a seed.  
Such seeds are often used by game developers, to generate the landscape and keep it consistent between sessions.
Our solution was to save the design as \textit{gene} expression that can be used to define the whole complex visualisation, and to enclose the seed in the gene expression.

Our fourth requirement is to enable designs to be \textbf{similar, yet different}. This helps to develop a range of designs, help to maintain brand identity across ideas and maintain a recognisable style and personalisation. For instance, using graphical elements to represent a collective group (e.g., a ward of patients) while ensuring each individual patient stands out visually. Or to enable users to create a variety of visualisation designs that share common traits yet are distinct and visibly unique. It should be adaptable and extensible, capable of designing both connected and non-connected visualisation types across a wide range, and apply to each part: object or colour palettes, paths, data and so on. We achieve this by our modular approach of using a \textit{path}, \textit{envelope}, \textit{object placement} and \textit{filters} and storing it in a \textit{gene}.

Fifthly, in addition to being deterministic, we require the visualisation generation algorithms to be suitable for real-time use and \textbf{computable}. Whether heuristic or exact decisions are made, we needed to be sure that we could compute and finish the design. The design constraints that we impose on our model are not as strict as, for example, with graph drawing~\cite{Battista:1998:GDA:551884} or label placement algorithms (e.g.,~\cite{Schulze_ETAL2018}). But there may be ways to make the model render depend on its context. Much like L-systems or context-grammars. Indeed our goal is to foster the creation of creative outputs, contrary to the goals of a developer who, when it comes to graph layout and label placement, is typically concerned with readability. In our case it may not matter if the objects overlap, and in fact, it may result in a nicer visual if they do overlap. In addition, we do need the algorithm to complete in real-time, and there may be situations where a heuristic is used to make a choice over two positions. We wish to avoid situations whereby the algorithm merely alternates between two choices and does not terminate. This is more an implementation challenge, and our solution is to focus on greedy approaches, such as to choose the first/ naive choice, or to follow a worsening choice strategy, like the Tabu search~\cite{Yamamoto_ETAL2002} algorithm. Finally, when such decisions are made, we need to save the seed or choice, such that the design is still deterministic.

Sixth, we require an \textbf{extensible} model, so that different path types, or layout algorithms may be used in the future. 
Our solution is to divide the model into modules: path, envelope, object placement and effect filters.
We believe that the path model is implicitly extensible, because of its composable nature; it would be easily possible to change the path to 3D, or add different elements onto the path, and to merge, clip or perform other functional operations to extend our basic ideas. 
\color{black}

\section{Related Work}
\label{SEC:RelatedWork}
Inspired by the structures of Reilly~\cite{Matesi2006}, we define a flowpath (line of action) and envelops (rhythms) to define a skeleton on which visual shapes can then be added to create a visualisation.  
This pathway represents the primary \textbf{visual flow} of the visualisation.  These observations align closely with theories of visual flow~\cite{pang2016directing,bylinskii2015eye}, which have been substantiated by eye-tracking studies~\cite{NielsonPernice2009}. We further discuss visual flow in the design goals in~\cref{SEC:DesignGoals} when we expand on our goals, and in~\cref{SEC:model} when we talk about paths. Ware~\cite{Ware2004} discusses a similar concept to reading a chart, often referred to as ``eyeballing the graph''. This concept emphasises the intuitive process by which individuals visually interpret graphical data, such as scanning a chart from left to right along the x-axis. Conceptually, this flow provides a convenient sequence, which we use as a backbone to create a visual design. Bertin's classification of image space usage~\cite{bertin1981graphics} uses icons to show the specific structures (linear, circular and so on) following also this idea of flow. 

Creating glyphs~\cite{borgo2013glyph} aligns with some of our goals as they provide a method to present complex data effectively, are scalable, and researchers aim to create variations that are both similar and distinctive.
We direct the reader to several relevant surveys. Ward~\cite{Ward2002} investigates glyph placement strategies. Borgo et al.~\cite{borgo2013glyph} reviews design guidelines, techniques and algorithms for glyph generation.  Ropinski, Oeltze and Preim~\cite{ROPINSKI2011392} survey the use of glyph visualisations for spatial multivariate data. Fuchs et al.~\cite{Fuchs2017} review 64 user studies of data glyphs, provide a meta analysis of the results and look into related design trade-offs. 
There are many ways to design glyphs (see Borgo et al.~\cite{borgo2013glyph}). For instance, 
Legg et al.~\cite{Legg2017} uses a quasi-Hamming distance to ensure sufficient individuality in the resulting glyphs. Ebert et al.~\cite{EBERT2000375} explore procedural generation, based on the automatic generation of glyph shapes using superquadric functions (also see Patel and Laidlow~\cite{patel2020visualization} and Gerrits et al.\ \cite{gerrits2019towards}), fractal surface displacement, and implicit surfaces.  Another strategy is metaphoric; to either take an object (such as a pea pod, as used by Koc et al \cite{koc2022peaglyph}) or  analyse an image to generate the component structures, that can be built into the final glyph (as used in MetaGlyph~\cite{LuETALMetaphoricGlyph2023} or glyph from icon~\cite{PresnovKolb2022}). An alternative approach is to evolve generations (e.g.,~\cite{lopes2020adea}) or use learning systems. Indeed, glyph learning systems is a dynamic and promising research area~\cite{LiuETAL2022LearningGlyphShape,Park2020}. A further strategy is to have an underpinning structure, to base the glyph on. For instance, Ying et al.\ \cite{YingETAL2022} in their circular glyph maker, define interior, intermediate and exterior structures. Brehmer et al.\ \cite{BrehmerETALDiatoms2022} use a generative approach, mixing a scaffold of a shape (rectangle, triangle, etc.), encoding channel and mark shape palette; similarly, Khawatmi et al.\ \cite{KhawatmiETAL2022} start off with simple geometric shapes. While Keck and Engeln~\cite{keck2022sparkle} use a star.
We, however, use a different structure. We use a flowpath as the backbone of the design. Our paths can be created manually, randomly, procedurally or from data (see \cref{SEC:model}). While Pereira et al.\ \cite{pereira2019generative} do use a simple skeleton line, their purpose is to create fonts, not glyphs or visualisations.

Glyphs have been used in a wide range of situations. For example,
Kindlmann and Westin explore glyph packing in medical visualisation~\cite{Kindlmann06}, whereas Maguire et al.~\cite{Maguire2012} propose a taxonomy-based glyph design approach, intended for visualising experimental design workflows encountered in biology.
Legg et al.~\cite{Legg2012} explore their use in real-time sports analysis.
Pearlman, Rheingans and Des Jardins~\cite{Pearlman07} focus on glyph-based multi-variate visualisation in different scenarios, such a analysing student applicant pools, network traffic and fantasy football team building. Kammer et al.\ \cite{KammerETAL2020Glyphboard} create a small-multiple zoomable dashboard of glyphs; in a similar grid-based approach Rees et al.\ \cite{ReesETAL2021} present hierarchical glyphs, to overcome the overplotting challenge when placing glyphs, and Pires et al.\ \cite{pires2020summarization} a treemap of glyphs. Suschnigg et al.\ \cite{suschnigg2021visual} use glyphs to present anomolies in time series data.
Glyphs are also used in more unconventional depictions, such as haptic glyphs~\cite{RobertsFranklin2005}, or in mobile Augmented Reality environments (e.g., Chen et al.\ \cite{marvist19}), or hand crafted for scientific visualisation~\cite{zeller2022affective}. And have been used to represent uncertainty~\cite{gerrits2019towards,fernstad2021explore}.
Many researchers have investigated the effectiveness of glyphs. For example, Maguire et al.~\cite{MaguireMotifs13}  investigate how motifs can increase  comprehension of workflow visualisations. While Hu et al.\ \cite{HuETALStarGlyphOrdering2021} analyse statistical orderings of star glyphs.  Other researchers have evaluated specific glyphs, such as whether Chernoff faces~\cite{chernoff1973use} are readable~\cite{naveh1982effect}, the effectiveness of temporal glyph designs~\cite{Fuchs:2013}, glyph placement strategies on maps~\cite{McNabbLaramee2019GlyphMaps}, analyses of historic use of music glyphs~\cite{nunez2019glyph}, effectiveness of flower- versus star glyphs~\cite{vanOnzenoodtETAL2022FlowerVsStarGlyph}, and how some designs are more glanceable than others~\cite{BlascheckETAL2018}.

Creating small visualisations aligns with some of our goals including rendering in a viewport, scalable, and quick to create. Tufte's~\cite{tufte2001visual} sparklines are small, simple, word-sized graphics with high resolution. 
Saito et al.\cite{Saito2005} use two-tone pseudo colours for quick data overviews, Willet et al.\ \cite{Willet2007} developed scented widgets embedding visualisations into UI elements, Perin et al.\cite{Perin2013} created sportlines for player movement in soccer, and Goffin et al.\cite{Goffin2014} designed word-scale visualisations with flexible placement. Brandes et al.\ \cite{Brandes2013} introduced Gestaltlines, multivariate sparklines using Gestalt principles. For a review of word-sized visualisations, see Beck and Weiskopf~\cite{Beck2017}.

Many developers have created tools to assist in visualisation creation. Early Modular Visualisation Environments (MVEs) like AVS, IBM DX, IRIS Explorer or Visage~\cite{SchroederETAL1996}, enabled users to craft custom solutions by connecting various modules together. Visual authoring systems like iVisDesigner~\cite{iVisDesigner14}, Lyra~\cite{ArvindHeerLyra2014}, and DataIllustrator~\cite{Zhicheng-DataIllustrator2018} assist designers in creating custom visualisations. While effective for bespoke designs, our focus is on rapid design and the generation of multiple diverse visual depictions.
Several tools assist in hierarchy visualisations: Visception~\cite{Yngve2020Visception} focuses on nesting, HiVe~\cite{Slingsby2009} uses a grammar for hierarchical layouts, and GoTreeScape~\cite{LiYuan2023} employs a declarative grammar. Preset-based visualisations~\cite{Schulz2015} and StructGraphics~\cite{Tsandilas2021} enable designing visual structures before applying data. Structured mock-ups rely on display space partitioning~\cite{Vuillemot2018}. 
Li et al.\ \cite{li2015exploring} identify features of visualisation designs that they explain as being similar to rRNA genes, that they use to build a high-dimensional phylogenetic tree.
Influential studies include Brehmer et al.'s mobile time range visualisations~\cite{Brehmer2018} and Blascheck et al.'s smartwatch glanceable visualisations~\cite{BlascheckETAL2018}, which we replicate using our path model (see case studies, \cref{SEC:case_studies}). Our method is similar to TimeSplines~\cite{OffenwangerETAL2024}, but allows path creation and attribute swapping with a drag-and-drop interface.

\begin{figure}[h]
    \centering
    \includegraphics[alt={Schematic of three examples, showing a simple path, path with jump, and path with several jumps.},width=\columnwidth]{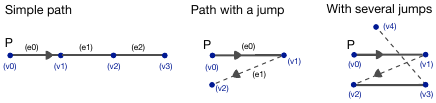}
    \caption{The path is made from a set of vertices. Objects encode the data and are mapped along the path.}
    \label{fig:ModelDifferentPaths}
\end{figure}

\section{The path-based model}
\label{SEC:model}
Referring to the simple example \cref{fig:flowgraph}, we define our path $P$ to be a directed path that is made from a finite series of vertices (see \cref{fig:ModelDifferentPaths}). While typically we equate vertices with (x,y) locations, it is possible to imagine that they could be mapped to other modalities; e.g., with sonification the sequence (path) could be notes on a music scale, sounds over time. A walk along this path therefore moves alternating between vertices and edges, $v_0,e_0,v_1,e_1,v_2,\ldots, v_{k-1},e_{k-1},v_k,$. Our path is a \textit{flowgraph}, in that we utilise every point in the path, and provide an order on the set of points from the first ($v_0$) to the last ($v_k$), which define the entrance and exit points, respectively. The path acts as a skeleton, on which we place visual marks. The path is contiguous, where there is a neighbourly structure from one vertex to another, however it does not need to be continuous, because we allow edges to jump. The path can also intersect and cross over itself, and can return to a previous position. However, this is achieved by adding the same points to the the path. The path itself does not hold any logic, it is merely a route to direct how the geometry is placed. Logical operations take place when the objects are placed along the path.

The flowgraph is used to express a \textit{visual flow} through a visualisation design space~\cite{eriksen1979information}. The way someone `eyeballs' a chart can be influenced by the cues surrounding it. The arrow of an x-axis and the nature that it is a line, draws the eye from the left side to the right. Visual designers can change how people look at certain parts of the picture; they provide visual cues to lead the viewer to \textit{flow} in a particular \textit{visual direction}~\cite{arnheim1965art}.  For instance, gradient lines direct the viewer along the direction of the lines, colour gradually getting darker would encourage the user to look in that direction, whereas size changes would encourage attention, while visual cues such as arrows would focus the observer to focus on the arrow head or whatever the arrow points to. The way we look at the visualisation is not random. We direct our eyes to interesting features, and follow the story of the graphic.

Visual flow is evidenced in many activities that involve a human looking through their eyes. For instance, a person reading an English text would read from the top-left of a page along a line to the right, and from the top of the page to the bottom (in a \textbf{Z-pattern}). Or someone looking at a webpage would glance at the top left and move down the page to the right. This dominance is named the \textbf{Gutenberg pattern}~\cite{pang2016directing}. The visual flow, allows users to eyeball the visualisation~\cite{Tufte1997} and permit the user to notice patterns such as heights of graphs, trends and outliers. Consequently, it is possible to analyse the visual flow and the dominant flow from the composition of the visualisation. In fact eye tracking can help to reveal these patterns of user behaviour, demonstrating for instance that web-users fixate on webpages in an \textbf{F-pattern}~\cite{NielsonPernice2009}. The F-pattern also suggests that most users will not read to the end of the row, but they skim down the page reading quickly and scanning the text. Eye tracking has been used widely in visualisation, for node-link diagrams, trees, and to understand how a user searches, perceives and explores a visualisation~\cite{bylinskii2015eye}. Matzen et al.\ \cite{matzen2017patterns} explain that visualisations are read differently depending on the task, and the type of visualisation, with unfamiliar visualisations inciting more eye tracked fixations. From these fixations they have developed salience metrics to predict how users will view visualisations~\cite{matzen2018data}. But even before eye-tracking studies,
Bertin~\cite{bertin1981graphics} classified several arrangements of how to utilise the image space, and provided a visual naming scheme based on these methods, including the
dimension of the plane 
\includegraphics[alt={Inline symbol of a right arrow},height=\mylen]{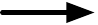},
repeated categories 
\includegraphics[alt={Inline symbol of a dotted right arrow},height=\mylen]{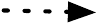},
collection of items
\includegraphics[alt={Inline symbol of a xn on a right arrow},height=\mylen]{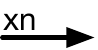},
cumulative quantities on a dimension
\includegraphics[alt={Inline symbol of a crossed out right arrow},height=\mylen]{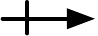},
circular 
\includegraphics[alt={Inline symbol of a circular arrow},height=\mylen]{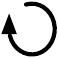},
irregular
\includegraphics[alt={Inline symbol of an S with an arrow head on its curved end},height=\mylen]{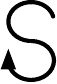},
and regular arrangement
\includegraphics[alt={Inline symbol of an S with arrow pointing right},height=\mylen]{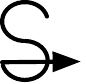}. These display models are summarised by Roberts~\cite{Roberts2000DisplayModels}.
Brehmer et al. \cite{BrehmerETAL2017} extend Bertin's scheme which they use to classify storytelling and timeline techniques.  There are similarities in their design space to ours; while they focus on specific instances, we have developed a path model that can be used to create different visualisations, and our gene allows us to easily change between several design strategies.
Additionally other path types are useful for our designs, including spiral~\includegraphics[alt={Inline symbol of a spiral, starting on the left and spriraling in},height=\mylen]{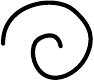}
and 
golden spiral
\includegraphics[alt={Inline symbol of a golden spiral},height=\mylen]{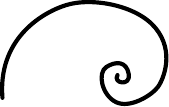}. By defining the vertices of the path we can recreate each of these styles.

\begin{figure}
    \centering
    \includegraphics[alt={Schematic of six sub figures, showing how the envelope can change or alternate how the visual object appears. Perhaps above, or below the path, or a mixture. The six schematics are: full envelope, only top envelope, only bottom envelope, decision per edge, z-pattern, z-pattern that switches the envelope.},width=.98\columnwidth]{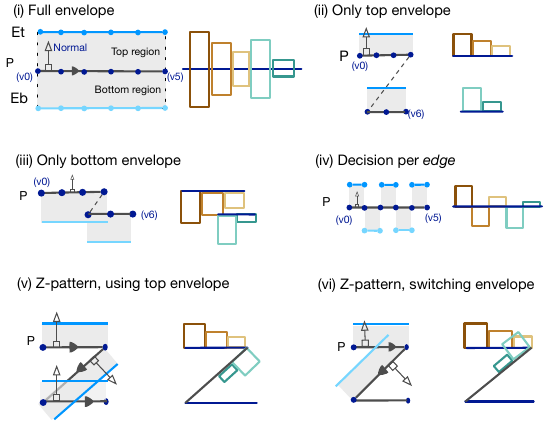}
    \caption{It is possible to control the placement of the objects by changing parameters of path and envelope. (i) Shows centred objects along the path; (ii) uses only the top envelope and snapping the objects to the path, (iii) puts objects under the path, (iv) can alternate between top and bottom envelop, (v) z-pattern showing how the return angle places objects underneath, (vi) and by switching the envelope on the angle objects can be placed above the path.}
    \vspace{-4mm}
    \label{fig:envelope}
\end{figure}

Our path is piecewise linear. We treat each of the edges as discrete parts. This is a sensible solution, because (i) the data that we map onto the visualisations will always be discrete in nature, (ii) it is clear how to map data onto the path, because it can be easily mapped along the edges, and (iii) even if the original data is continuous in nature, a developer we would have stored a discrete (sampled) version of it in a database.  This strategy does not inhibit continuous or smooth-looking visualisations; on the contrary we control such smoothing at the envelope (\cref{sec:Envelope}) and object placement stages (\cref{sec:objectPlacement}). We note that it would be possible for us to have defined a curved path, such as the cubic beziers definition in the SVG PATH. Yet, one disadvantage with this solution is that while such a path would be continuous it would be unclear how to map the data onto it. 

Finally, it would be possible to create unsuitable paths, such as those that are unnecessarily complex, contain too many vertices that may take a long time to render, or paths with the points that are all located in a small space that produce overplotted images. However, our goal is not to constrain potential creativity, and hence we do not constrain the length or position of the vertices in the path. Nevertheless, in our implementation we make checks on the path, e.g., eliminate null vertices, that the smallest object is larger than a pixel, and clipped to render in a unit space.  

\subsection{Envelope}
\label{sec:Envelope}
The \textit{envelope} is defined by a finite set of vertices of a top ($E_t$) and bottom ($E_b$) path, and sits on the main path ($P$). The envelope has a Normal ($N$) that defines a vector perpendicular to the path $P$, which is used to determine the top and bottom envelope regions. While the \textit{path} describes the visual flow and positioning of the elements, the \textit{envelope} determines how the objects appear along the path, and defines the baseline of where the objects are placed. For example, the objects can be placed \textit{centred}, \textit{above} or \textit{below} the line through controlling which part of the envelope is used, as shown in \cref{fig:envelope}. The envelope defines a range, which is the viewing region where the graphical objects will be drawn. It can be used to clip the visualisation object should it be larger than the range of the envelope, or filters can be applied (during the object placement stage) such to smooth or filter and merge objects together, see \cref{fig:envelopeOverlap} and \cref{sec:objectPlacement}.

\begin{figure}[h]
    \centering
    \includegraphics[alt={Two schematic diagrams. The first shows a circle with rectangles on the inner path, representing a bar chart styled visualisation on a circle. The second a spiral bar chart.},width=1\columnwidth]{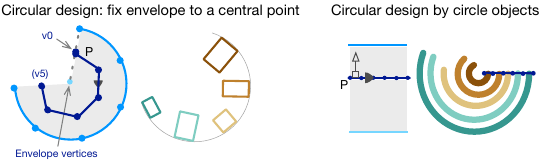}
    \caption{Different circular visualisations can be created through circular paths, and by setting the envelope points to a fixed location. Or by placing circular objects.}
    \label{fig:Circular}
\end{figure}

In many situations it is easy to conceive that the envelope $E_t$ and $E_b$ will be parallel to the path. The envelope is defined by a series of vertices (much like the path), and progresses along with the path to define the visual flow of the visualisation. \cref{fig:envelope} shows some example path and envelope combinations. \cref{fig:envelope}(i) demonstrates the full envelope, with a parallel path, which is used to centre the objects as they are placed along the path. Not only can the path be non-continuous, but also we can determine if the objects sit above or below the path, by changing the envelope, as shown in \cref{fig:envelope}(ii) and (iii). In fact, we can use this technique to create a small-multiples display. \cref{fig:envelope}(iii) also shows that it is possible to create overlapping objects. We can switch the choice of the envelope, such to move the baseline of the object from top to bottom, as shown in \cref{fig:envelope}(iv).  Finally, \cref{fig:envelope}(v) and (vi) show zig-zag paths, where (v) is the normal situation whereby the path naturally turns the objects upside down, as it goes through the diagonal part of the Z-path. This example is important, because a designer may wish to keep the objects upright. This is achieved by switching between top and bottom envelope, when the path turns.

By changing how the path is constructed and how the envelope and path work together, and how the objects are placed we can describe different structures. In particular, the envelope does not necessarily have to follow the path, so for instance, by fixing all the vertices of one envelope ($E_b$ in this case), we can create a circular visualisation (\cref{fig:Circular}). An alternative circular visualisation design can be created using
a linear path, and placing circular objects.

\subsection{Object Placement}
\label{sec:objectPlacement}

\begin{figure}[h]
    \centering
    \includegraphics[alt={Five small pictures, showing a fish-like shape overlapping with a smaller yet similar object, which are subtracted, to create two stream-like shapes.},width=\columnwidth]{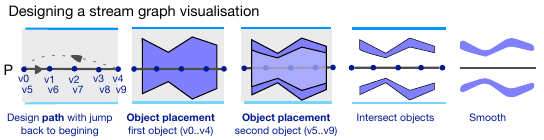}
    \caption{Building up a stream graph visualisation, through several coincident paths.}
    \label{fig:streamgraph}
\end{figure}

Once the \textit{path} and \textit{envelope} have been defined, we now decide what \textit{objects} to place down, and how to apply them. These objects are the geometrical shapes that become the visual marks that encode the data. We have used many shapes including circles, arcs, donuts, triangles, rectangles and lines. The location of the objects is controlled by the path and envelope, but the visual appearance and any effects are applied in this step. Objects are usually placed in path order and overlapped, where subsequent objects will be placed on top of previous ones. Therefore, if a different order is required, either the path vertices need to be re-ordered, or the data order changed. When the objects are placed they are positioned, scaled, rotated, and any filters applied, before applying the clipping of the path, or other constraints applied by the path and envelope. For example, a stream graph visualisation can be created by defining a path ($P$) drawing one outline, using the same vertices to draw the second outline, subtracting the objects, and smoothing it with a filter (see \cref{fig:streamgraph}).

\begin{figure}[t]
    \centering
    \includegraphics[alt={Four smaller sub-figures of a line of circles. Shows combined circles to make a smooth metaball shape.},width=\columnwidth]{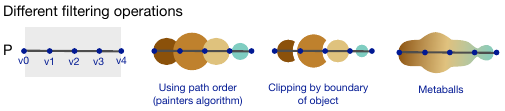}
    \caption{Illustration of an envelope overlap and three possible strategies for dealing with it: 1 objects are placed in order, with the next object overlapping the previous, 2: The previous object is cutout, 2: The objects are combined using a function (in this case Metaballs).}
    \label{fig:envelopeOverlap}
\end{figure}

By placing different objects and applying a variety of filtering operations we are able to create typical visualisations, as well as new designs. There are many filters that could be  applied, including smoothing, intersection, union, blend, blur and shadow. Overlapping objects allow several effects to be applied. \cref{fig:envelopeOverlap} shows three scenarios: (i) objects placed on order, with later objects being positioned on top, (ii) later objects cutout sections, and (iii) combining objects with a metaball operation~\cite{BlinnMetaBalls0}.

\begin{figure}[t]
    \centering
    \includegraphics[alt={Four subfigures. The first three present how the scatterplot is created by moving along a path and drawing objects. The last subfigure shows a set of fourteen points that are connected by an angled line, to demonstrate the scatterplot being created in data order.},width=\columnwidth]{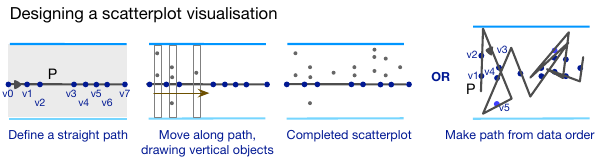}
    \caption{Two strategies for representing a scatterplot using the path model. 1) The path is defined across the design space and scatterplot marks are placed vertically. 2) A path is constructed from the data in order and scatterplot marks are placed along that path.}
    \label{fig:scatterPlot}
\end{figure}

\begin{figure}[t]
    \centering
    \includegraphics[alt={Two subfigures. The top with eight small graphics, showing different path types with the first two as follows: named `inline'  it shows a typical bar chart, named `disjointed linear' it shows a picture of a barchart split over two lines. The second line of small images presents nine different small pictures of space-filling curves. The first three are as follows: named sweep, is a z pattern. Named `scan' is a z pattern that is squared off. Named `diagonal' is a z pattern on an angle.},width=\columnwidth]{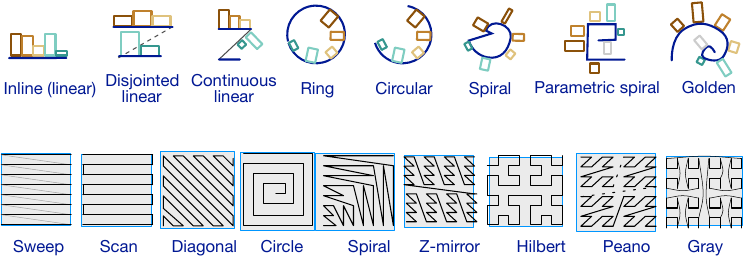}
    \caption{Example path types. Lower are space-filling path types, including sweep, scan, diagonal and Hilbert, Z-mirror, Peano, and Gray~\cite{connor2010fast}. Each space-filling path traverses every grid point, in a continuous, non-overlapping manner. Hilbert, Peano, Z and Gray look to place nearby points in the grid such they become close together in the curve.}
    \label{fig:differntCurves}
\end{figure}

We can define more complex visualisations using more abstract paths, see~\cref{fig:scatterPlot,fig:differntCurves}. For example, to create a scatterplot (\cref{fig:scatterPlot}) either the path is defined as a straight line across the $x$ axis, where the vertices are determined by each of the $x$ data values, and the objects being placed are the scatterplot circle symbols. Or an abstract path is defined, that joins every datapoint together, and is positioned such that each data point sits on the \textit{edge} between two vertices. The scatterplot is drawn by moving along this abstract path. This second approach would be helpful if considering label placement. Other paths can be implemented,

\vspace{3mm}
\subsection{Data mapping}
The data mapping stage of our model offers some interesting choices and opportunities.
We can map \textit{data} to any part of the model, such as the \textit{path}, \textit{envelope}, object, object position and its appearance, and apply filter operations on the objects (based on data).  For example, a simple bar chart is created by defining a regular set of path points ($v_0$ to $v_{n}$), choosing an envelope that sits on the path, placing rectangles such that they sit on the path, and mapping each data value to adapt the height of the rectangle. This will create a black set of rectangles (the default colour), so to add different hues we can map a colour to the fill of each rectangle. We could adapt this further, for example by mapping data to vertex position we can change the width of each rectangle. Same data values could be simultaneously mapped to different parts of the model, or data to different parts to realise a multivariate visualisation.

Data mappings can be dependent on previous decisions. This is possible because we have a precise ordering to our model (path creation, envelope, object placement and mapping).
For instance, to create stacked bar charts, subsequent rectangles start where the previous ends. Trees and other hierarchical structures can be similarly created. While we do not allow conditional logic to be embedded in the path, we do allow the path to jump, therefore simple tree structure can be created by placing an object, and then jumping the path to the start of that object. While the path does provide a skeleton, and describe the general visual flow, (in the most extreme cases) the geometric objects can move away from the physical constraints of the path. For instance, a rectangle can be positioned on an edge, with its exact location being jittered (or offset from a data value). However, when such operations are applied, it is important to consider the requirements of being composable, self-contained and deterministic. Any seed value (for stochastic jittering) or data values for relative positioning, must be encoded into the gene expression such make certain the visualisation is deterministic.

\vspace{1mm}
\section{Implementing the path model}
\label{SEC:implementation}
\begin{figure}
    \centering
    \includegraphics[alt={A grid of 24 visualisations, demonstrating the range of output from the Genii system. They represent colourful blocks, intersecting objects, placed in a variety of circular, zigzag and different arrangements.},width=\columnwidth]{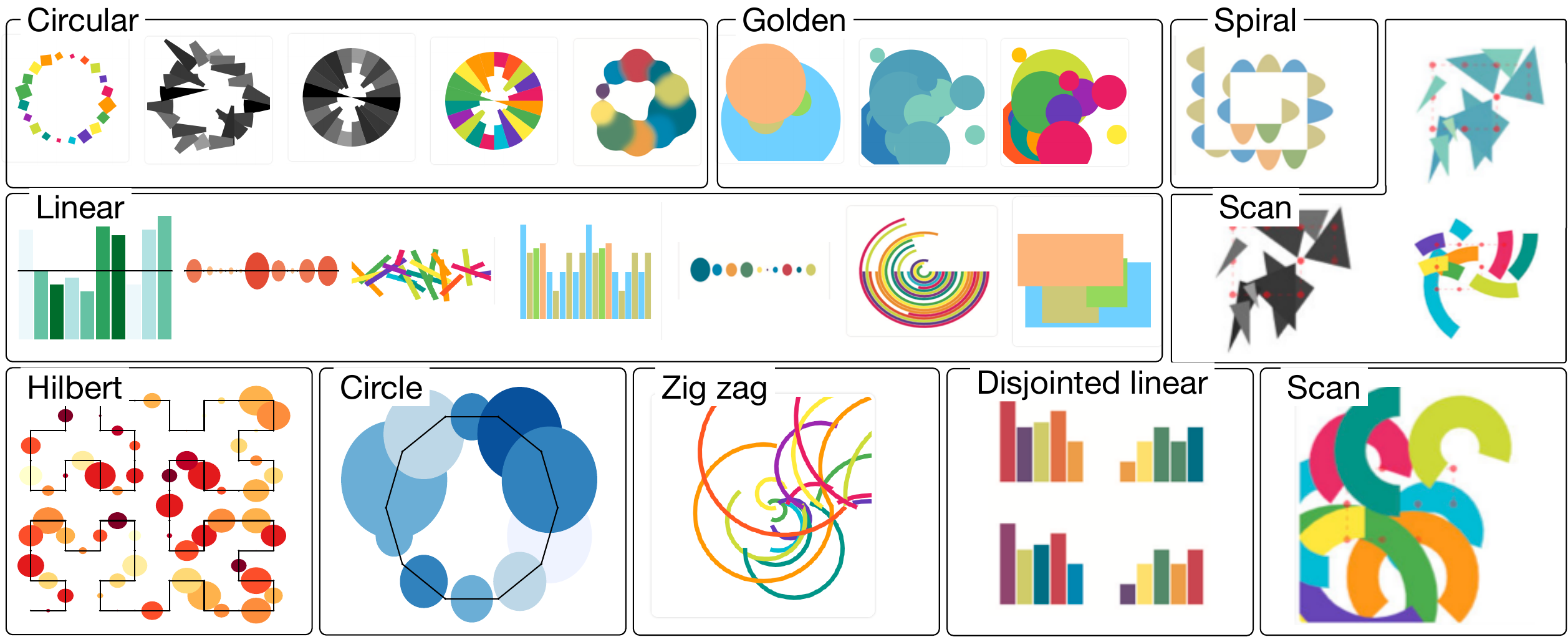}
    \caption{Genii visualisation designs, showing a range of  different flow paths. }
    \vspace{-4mm}
    \label{FIG:GeniiPict}
\end{figure}

The Genii builder is shown in~\cref{fig:teaser}, resultant output in \cref{FIG:GeniiPict}, and case studies in~\cref{fig:caseStudy}. The system consists of a component library, renderer and builder application, see  Supplemental Materials. 
We first made sketches, which helped us to refine the initial concepts and the model. We implemented an initial prototype with JavaScript and HTML5 Canvas~\cite{Jackson-et-al-Poster-VIS2018}. This prototype included features for object unions, intersections, and metaballs~\cite{BlinnMetaBalls0}. However, we encountered difficulties  in creating paths that depended on the positions of other objects. 
We needed a more modular approach to the implementation. We chose React.js because it is modular in approach, uses HTML5, CSS and JavaScript, and in particular we are able to pass detailed data to control the DOM. It also provides rapid prototyping and multi-platform support. It also allowed us to develop the full browser application and migrate it to the Apple watch, which links with a iPhone, see~\cref{fig:mobileTool}. We detail the implementation below:

\begin{figure}[b]
\centering
  \includegraphics[alt={Photograph of a person's arm. They are wearing an iWatch on their wrist. The iWatch screen shows a circular barchart. Alongside is a smart phone (iPhone), also running the Genii system, showing the same circular barchart.},trim={0 32cm 0 20cm},clip,width=\columnwidth]{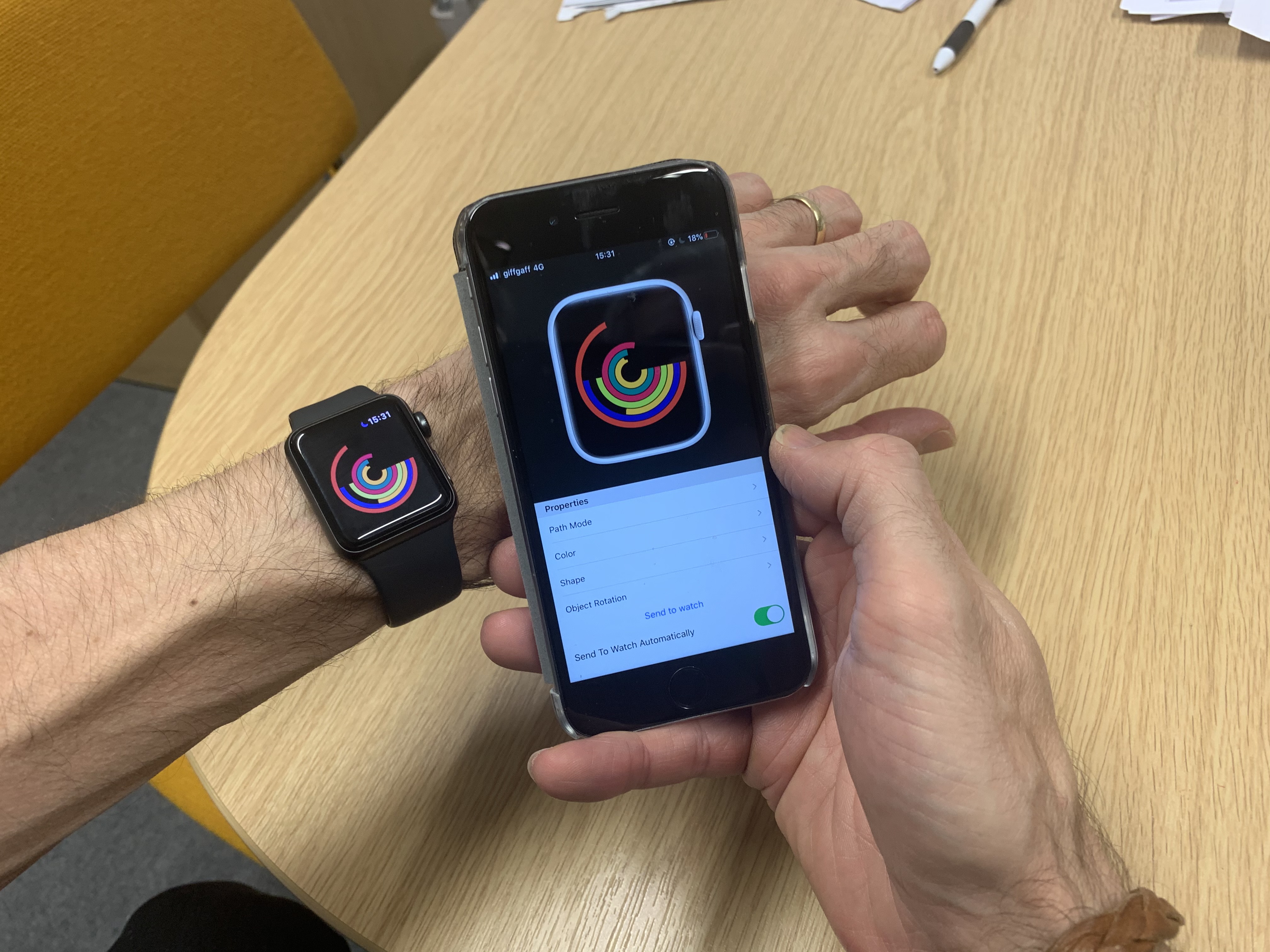}
  \caption{Photograph of the Genii iOS application where people can build the visualisations and load them to the companion smart watch. %
  \label{fig:mobileTool}}
\end{figure}

\textbf{Flowgraph}. There are three ways to implement a path in Genii. First we define set paths. These include: \textit{inline linear} which provides a uniform  straight line distribution of points horizontally across the canvas. \textit{Disjoint inline} a path with jumps. \textit{Ring} which is a circle of points, with the same first and last coordinates. \textit{parametric spiral} which is formed by a series of straight line paths with $90^o$ rotations and increasing path lengths. Hilbert space filling curve that passes through every point within a square grid in an uninterrupted, non-overlapping fashion. Each path is adaptable, for instance, by anchoring one reference point to the center of a circular path, we can generate a star plot. Second, we provide a path editor for user-generated paths. Individuals can visually place points, and Genii records the path accordingly. Finally, paths can be added to the code, by replicating and editing the React component.

We implement the \textbf{envelope} in Genii by clipping. Elements are placed (centred aligned, top or bottom aligned) along the path, based on the Normal.  Which can be used to create objects that alternate between above, below or on the path.
\textbf{Object placement} is achieved by placing geometric shapes between vertices of the path. As shown in~\cref{fig:scatterPlot}, the distance between vertices determines the default location of the object placement. Then the data values will then adapt this position. For example, one chosen shape can be mapped along the path, and its location adapted by the data. Or the shape can alternate its placement (sitting on the line, or below it) dependent on its position. Or different shapes can be applied, to each location along the path, dependent on the data. 
Genii includes several basic geometric shapes, including: \textit{lines} that can be used to create triangles, or outline shapes; \textit{triangles} by drawing a line between the vertices and the central point is adjusted by the data value; \textit{arcs} using the central as a B\'{e}zier control point of a quadratic curve, rectangles; \textit{circles} and \textit{ellipses}, where the radius can be controlled by the distance between the vertices, or by a constant, or by a data value. We have implemented several filter operations, including object/primitive fill, linear gradient fill and radial fill, change of object stroke, and metaball merge. 

We needed a way to load  \textbf{data} and define the sizes of the visualisation. We store the data in a JSON object that includes the name (a string), data value (float) and data range(float) of the data for each category, the \textit{width} and \textit{height} of the visualisation (in centimetres), and padding around the edge of the visualisation. This provides a convenient way for us to save the data created in Genii.
We store the visualisation design in our\textit{gene} expression. Users hover over the top right hand side of the Genes panel (see~\cref{fig:teaser} middle). The saved file contains a set of enumerated types stored as parameters in JSON.  Our application will only save valid genes, but should they be invalid (edited by an external user) our application first validates the gene expression and warns the user where the problem lies, this is achieved using the transpiler (Babel.js). The gene saves many properties (see~\cref{lst:listing2}), including: Path Mode, Path Rotation, Path Point Distance, Path points, Object Shape, Object Colour, Path Grouping. To make the application \textit{deterministic}, the user is asked to \textit{name} their visualisation, which is hashed to a 32bit integer value. We use a hashCode() function of string $s$ to create the product sum of the string's character codes, as follows $h(s) = \sum\nolimits_{i=0}^{n-1}s[i]\cdot 31^{n-1-i}$.\\[-0.5em]

\lstset{numbers=left,xleftmargin=3em,framexleftmargin=2em,basicstyle=\small}
\begin{lstlisting}[caption=Example gene; used to save the path-based visualisation, label=lst:listing2]
const params = {
  shape: Gene.shape.BAR,
  color: Gene.color.FROM_DATA,
  color_key: Gene.color_key.OFF,
  path_points: Gene.path_points.EVEN,
  path_mode: Gene.path_mode.INLINE,
  path_rotation: Gene.path_rotation.NONE,
  path_grouping: Gene.path_grouping.NONE,
  object_rotation: Gene.object_rotation.NONE,
  object_size: Gene.object_size.FULL,
  filter: Gene.filter.OFF,
  debugging: Gene.debugging.OFF
}
const Gene = new Gene(params);
\end{lstlisting}

\section{Case Studies}
\label{SEC:case_studies}
To demonstrate the model we recreate several visualisations from papers (Blascheck et al.~\cite{BlascheckETAL2018} and Brehmer et al.~\cite{Brehmer2018}) and an imaginative design from a concept sketch. Each of the visualisations from the papers are small visualisations that would be suitable for mobile devices. For each, we present their visualisation and analyse the structure, show a copy of their original visualisation and explain how we recreate them in our path model.

\begin{figure}[ht]
    \centering
    \includegraphics[alt={A grid of 30 smaller images, in 5 columns. The columns are a screenshot of the original visualisation, a schematic of the path, of the path with placed object, and with the object adjusted, and the last colunmn is the rendered visualisation. The first three rows recreate Blascheck et al. The next two recreate visualisations from Brehmer et al. The final row recreates a visualisation from sketch.},width=\columnwidth]{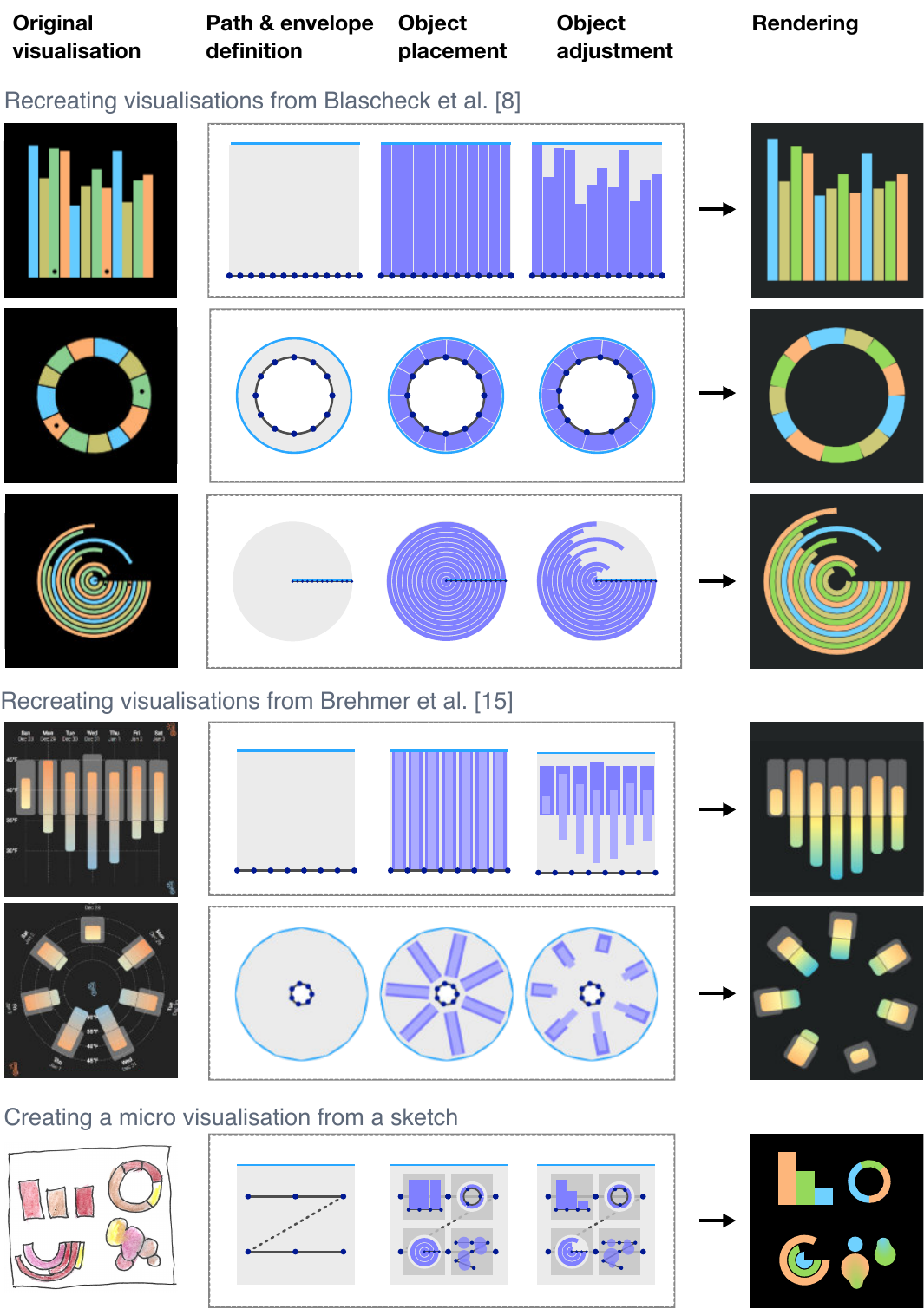}
    \caption{Recreating visualisations by Blascheck et al.\ \cite{BlascheckETAL2018} (top) and Brehmer et al.~\cite{Brehmer2018} (middle). Also creating a visualisation using a non linear path from a sketch. Showing the original visualisation, an explanatory depiction of the path used, and the final rendering (right).}
    \vspace{-2.5mm}
    \label{fig:caseStudy}
\end{figure}

To recreate the bar chart, donut chart and radial bar chart of 
Blascheck et al.~\cite{BlascheckETAL2018} we need to imagine the visual path.
The left column of \cref{fig:caseStudy} (top) shows their original visualisations, our paths and our recreation,
see \cref{fig:caseStudy}). 
Understanding the visual path of the bar chart is easy, where a straight path would readily recreate this visualisation. Similar to the examples in \cref{fig:envelope} we recreate this chart using a straight path, envelope that stretches to the top of the design space, rectangles aligned on the path encoded with data. The donut chart can be created in several ways. We could treat this as one circular object that is placed on a simple path (see \cref{fig:Circular}) or a circular path. Circular paths can differ, depending on whether the objects would sit at the top of the envelope or the objects sit on the path. The best decision depends on how the donut will be used, and how it displays data. 
Because the width of the donut's ring is not defined in the original paper we decided to create the donut chart with a circular path, truncated circular segments of a cone (the objects) sitting on the path. There are different ways to implement the radial bar chart (as discussed before, see~\cref{sec:Envelope} and \cref{fig:Circular}); we chose to implement it using a straight path with circular objects, encoding the data to fill a percentage of the full 360$^o$  arc. For each visualisation the colours follow a colourblind friendly differentiator scheme between data visualisation elements.

\setlength{\belowcaptionskip}{-5pt}

We recreate two visualisations from Brehmer et al.~\cite{Brehmer2018}, shown in \cref{fig:caseStudy}, which can be done in different ways.
For the first visualisation we use a straight path; but on the outset it may not be clear to ascertain how many paths are required. Either the \textit{flow} is considered as one path and we place a complex object at each point; or it can be appraised as several paths, that are located at the same point (like the streamgraph example, \cref{fig:streamgraph}). While the outcome would be the same, differences exist in the way to build the path and apply the data. 
We use a straight path with envelope above it. There are two other possible path and envelope combinations: i) path through the middle with envelope above and below; ii) path along top with envelope below. Alternative i) requires extra checks on the data and a more complex placement calculation. The differences between alternative ii) and the chosen path are subjective and we make the decision to place it at the bottom was made to match with an $y$ axis. In addition we map colour. Three visual variables are used to describe the data: a) colour is a linear gradient from blue to yellow to orange for daily temperature range, and an opaque white colour to denote the average temperature range; b) the size of the element shows the size of the range of the data point; c) the placement of the element shows the start of the range. We map a filter to the objects, to round the corners (in the same stylistic way), and we do not include labels (but could do so by adding a new path, and placing text labels). We create the circular visualisation using a similar path to the circle chart of Brehmer et al.~\cite{Brehmer2018}, but apply different shapes and align the objects to the outer part of the envelope.

Finally, we create a disjointed design which has four separate visualisations (see \cref{fig:caseStudy}). We sketched the idea and then created a disjointed path.  We start by defining z-pattern path, we insert extra vertices into each part of the z shape, such that we can describe the individual visualisations. For the first three (bar, donut and circular chart) we followed the descriptions  as described above. The final visualisation uses a random path, and place circles along the path, applying a metaball filter to blend the circles.  

\section{Browser and mobile app usability evaluation}
\label{sec:Evaluation}

Our emphasis lies in crafting diverse designs rather than assessing the effectiveness of the resulting visualisations. For evaluations of their efficacy, we direct readers to the work of other researchers in the field (e.g.,~\cite{BlascheckETAL2018,Brehmer2018}).
To evaluate the usability of our path model, we used our desktop implementation (see \cref{fig:teaser,FIG:GeniiPict}) and mobile phone/watch tool (see \cref{fig:mobileTool}), with  default paths available. Users were given a five minute introduction video showing how to use the tool, followed by five minutes self-guided exploration, where they  could create their own visualisation designs. They identified if they ``liked'' or ``disliked'' each creation during the desktop study by clicking on the images in the gallery. On mobile, users had the ability to send the visualisation to the Apple Watch if they liked the design. We chose a short design duration, because we wanted users to rapidly create many visual designs. After five minutes they completed the System Usability Scale (SUS)~\cite{brooke1996sus} questionnaire. Participants also reflected on their design process, by answering (1) ``Describe how you would like to use these visualisations in day to day life''; (2) ``Describe some aspects that you liked about the visualisation design process''; and (3) ``Describe some aspects that you disliked about the visualisation design process''.

\textbf{Participants.} An initial pilot study was carried out using four participants (two web developers, one psychology student and a student nurse) recruited over social media. We used our JavaScript  evaluation framework~\cite{Jackson-Roberts-VIS2017} that allows users to anonymously perform the evaluation. From this feedback we improved the gene encoding strategy (changing it from an initial drop-down menu to a drag-and-drop interface), added a colour for each gene, and made the gene clearer (through the drag-and-drop interface). For the browser study, 18 participants were recruited via social media covering a range of backgrounds.  Because of our anonymous system we do not know who completed our evaluation, but from conversations and where we advertised we conclude: eight were general public (not at the University) and four postgraduate and six undergraduate students (from a range of computing, nursing and psychology). Separately, the mobile study was conducted under controlled conditions using a single iPhone and Apple Watch, this study comprised two groups, the first was eight final year Computer Science students, the second group comprised nine members of the public.

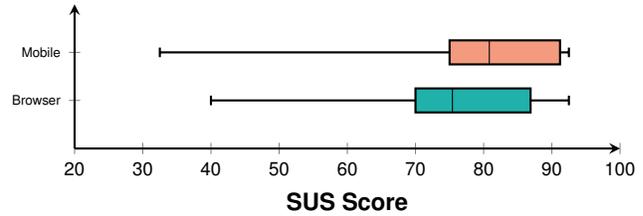
\begin{figure}[t]
	\begin{tikzpicture}[font=\sffamily\sansmath]
	\begin{axis}
	[boxplot/draw direction=x,
	xtick={10,20,30,40,50,60,70,80,90,100},
	xlabel={\textbf{SUS Score}},
	ytick={1,2},
	yticklabels={Browser, Mobile},
	y label style={},
	x label style={draw=none},
	x tick label style={draw=none, at={(axis description cs:0.1,-2)},rotate=0,anchor=north, font=\small\sffamily\sansmath},
	y tick label style={ font=\tiny\sffamily\sansmath},
	ymin=0, ymax=3,
	xmax=100,
	xmin=20,
	width=\columnwidth,
	height=3.5cm,
	axis x line=bottom,
	axis y line=left,
	xtick distance=1,
	axis line style = thick,
	]
	\addplot+[fill=myGray,
	boxplot prepared={
		median=75.42,
		upper quartile=86.88,
		lower quartile=70,
		upper whisker=92.5,
		lower whisker=40,
		draw position=1,
		box extend=.5,  
		whisker extend=0.2, 
		every box/.style={thick,draw=black,fill=editorTeal},
		every whisker/.style={black, thick},
		every median/.style={black, thin},
	},
	] coordinates {};
	\addplot+[fill=myGray,
	boxplot prepared={
		median=80.83,
		upper quartile=91.2,
		lower quartile=75,
		upper whisker=92.5,
		lower whisker=32.5,
		draw position=2,
		box extend=.5,  
		whisker extend=0.2, 
		every box/.style={thick,draw=black,fill=diagramOrange},
		every whisker/.style={black, thick},
		every median/.style={black, thin},
	},
	] coordinates {};
	
	\end{axis}
	\end{tikzpicture}
	\caption{Box plots of SUS scores. The usability of each tool was good, and therefore should not affect adversely the generation of each visualisation.}
	\vspace{-3mm}
	\label{fig:susscorebox}
\end{figure}

\textbf{Results.} In the browser study the SUS survey was fully completed by 15 of the 18 participants. We calculate the mean SUS score to be 75.42 (\cref{fig:susscorebox}) indicating a `good' score according to Bangor et al.~\cite{Bangor2009}. The minimum score was 40 while the maximum score was 92.5. The lowest score is interesting, because while they answered low values in the SUS each of their comments were extremely positive; perhaps they misunderstood how to complete the SUS Likert scale. For the mobile study all 17 participants completed the SUS survey. The mean score was 80.83 indicating a `good' score. The minimum 32.5 while the maximum was 92.5. The low score for the mobile study was accompanied by largely positive comments except that the labelling of properties did not make much sense to the user. 
Along with the SUS, we can analyse the created visualisations. 

\textbf{The browser study visualisations.} The 18 participants created a total of 163 visualisations. The minimum number of visualisations created was 4 and the maximum 15. This means that, on average, 9.06 visualisations were created per participant within five minutes.  The survey automatically time-stamps every action, and especially it automatically progresses to the SUS after five minutes. Therefore we are able to evaluate the duration of the design process. The average time spent designing was 356.5 seconds with a minimum of $302.2$ seconds and a maximum of $689.3$ seconds. Using this data we can infer that users of the model are able to create a visualisation every 39.37 seconds. Of the 163 visualisations that were created 89 were liked (through self judgement) and the remaining 74 disliked. 

\textbf{The mobile Study visualisations.} During the mobile study 19 participants created a total of 156 visualisations with a minimum number of 3 and a maximum of 12, this averages to $8.21$ visualisations per user. Within the given five minute period participants created, on average, one visualisation every $36.54$ seconds. 106 of the  visualisations were sent to the Apple Watch indicating that $67.9$\% of the visualisations created were liked.
The binary choice between \textit{like} and \textit{dislike} forced participants to make clear decisions over the rendered visualisation and whether amendments of the gene were needed. 157 of the visualisations in the browser study and 111 during the mobile were created using a unique combination of gene parameters, this adds credence to our assertion that the model allows users to create many different designs in a small amount of time. 

\setlength\belowcaptionskip{-5pt}

\section{Discussion and Conclusions}
We believe the path model has huge potential to create new visualisations, and by manipulating objects on the path by the envelope and filtering operations and storing the gene we are able to create a wide range of visual designs. 
In fact, participants in our evaluation said that the model was \textit{``simple to use''}, another said \textit{``very easy to use''}, and it \textit{``could create some interesting designs very fast''}. Indeed, our evaluation demonstrated that participants were rapidly creating visualisations, between 4 and 15 in five minutes. 
Some participants in our evaluation said that they would like to use it for \textit{``mobile or smart watch applications, maybe for health or fitness indications''} with another participant saying that it would \textit{``integrate into smart devices where you could actively quickly design and change backgrounds for devices''}. Other participants said that it would be \textit{``suited to a dashboard including other more detailed plots, or as a summary visualisation on a mobile application''}.

The concept of using the \textbf{visual flow} as the main underpinning structure for creating visualisations is a new way to contemplate visualisation design. 
While other researchers  have considered paths or skeleton constructs (e.g.,\cite{bertin1981graphics,BrehmerETAL2017}), they have done so to classify visualisations, or to analyse analysed the visual flow of pictures (e.g.,~\cite{pang2016directing,bylinskii2015eye}), whereas Pereira et al. utilise a simple path \cite{pereira2019generative}. However, our work places the \textit{flowgraph} as the core visual skeleton to design  visualisations. Like with the line-of-action lines of Reilly~\cite{Matesi2006} the flowgraph provides an excellent and understandable structure to underpin  visualisation designs. 
Integrating the flowgraph into visualisation design not only provides a structured framework but also enhances the creation of alternatives. By defining how elements flow and interact within a visualization, the flowgraph enables designers to create narratives, emphasize relationships, and guide viewer understanding intuitively. This approach has potential to support dynamic visual storytelling. Moreover, leveraging the flowgraph as a foundational design element encourages exploration and innovation in visual representation techniques, paving the way for new insights and discoveries.

One of the challenges for a designer, though, is that they need to think creatively, to first imagine what the path would be, and then decide on how to map the objects along the path.  Most visualisations are designed from the data up; where the visualisation designs are created pragmatically through mapping data to change visual variables that are placed on the display. However, in our approach, developers need to conceptualise the visual flow, and use this to structure their visual design (though, much like a sketch artist would do). From our  evaluation and case studies (especially) it is clear  that some paths are easy to comprehend. It is clear that a bar chart has a linear flow, it is clear that a pie chart would have a circular path; but for some designs the path will be less clear. What path would be best for a tree? What would be the best path to label points on the graph? Such complex paths can be approximated using multiple paths or several jumps in the path, but there may be other ways to achieve this. When planning or designing new visualisations the user needs to think about the design and iteratively refine their ideas, perhaps starting off by sketching designs~\cite{RobertsHeadRitsos2016} before implementing them. We believe a similar approach is needed here. Visual designs will emerge, and the developer needs to contemplate what the visual path will be. Sometimes the designer will create a path and realise that it was the wrong strategy. Developers will need to learn how to create the `best' and most efficient path.

\textbf{The gene} provides a quick way to re-factor designs. We demonstrated this in our second case study. Once the path, envelope, object placement, filters and data mapping have been created, it is easy to swap out some of the parts to create a new design. In that case, turning a straight path into a circular one. Such a strategy would be difficult with traditionally built visualisations.  The idea of the gene was positively discussed in our evaluation. Once participant said ``\textit{love how you can alter the final visualisation without having to start from scratch}''. Some did say that some genes do create similar visualisations, but we believe this is the result of their use of the tool and choices they made. There was some confusion over some of our name choices, for example one participant wrote:  \textit{``..I did not know what an `inline half' was before I created the visualisation and even though I liked the outcome''}. When utilising the path editor, users have the freedom to define their preferred path.

One limitation with Genii is that it is not always easy to change from one visual flow to another. For instance, swapping between a straight path to create a circular visualisation using circular objects, into one that uses a circular path with straight objects currently requires the user to re-define the path and swap the mapped objects. But similar challenges exist when programming these visualisations from a library, and refactoring the code. We are currently improving our implementation to allow this pivoting operation to be simpler.
There are other improvements we could make. One participant said \textit{``I would have liked a textual or visual description of what each property would do to my visualisation.''} and another \textit{``when there were many visualisations in the right hand panel, knowing which gene corresponded to which visualisation was tricky when I wanted to copy and modify them''}. We are working to develop and improve the implementation.

There could be several extensions, or further ideas to our flowgraph model, that we could explore. \textbf{Paths-in-path} could simplify the creation of hierarchical structures. We could define several paths, name them, and use them later. This re-use could make path design easier. However, we did not include path-in-path because our flowpath model already allow jumps, can contain infinite vertices, the path can depend on complex data, which in turn can be used to provide the same functionality as paths-in-paths would provide. Another issue would be that adding such path-references and paths-in-paths would make the input parser more complex, and it is not necessarily requires. It is already possible to create an application to output more complex paths that could encode several complex jumps to that would be necessary to model a tree, treemap or other hierarchical visualisation. 
We also considered adding \textbf{logic into the path} itself. For instance, we already make decisions on the path which are based on the data. This is used to alternate the position of the objects on, or above the path, to change their colour and so forth. But it could be possible to embed logical decisions on the path. To achieve this functionality we would need to add ID labels to control jumps within the path itself. This would  provide another way to build hierarchical (or other complex) structures. Applying logic to the path could allow element reuse, and recursion to occur within the design process. However, this would turn our path model from a scene description language into a programming language. From deterministic to non-deterministic. There has been a similar discussion within the scene graph community -- where researchers have discussed the placement of logical statements within the scene~\cite{Johnson_2015_CVPR, Xu_2017_CVPR}. Much like standard scene graph implementations~\cite{He_2016_CVPR, ren1506towards, Wang2009}, our model does not include logic at the path level. But not having logical, variables or ID's embedded with the path does not restrict complex paths to be created; we can still code a program to write more complex paths, which can then be used as before.

Another addition is to \textbf{add labels and legends}. One solution we contemplated was to add a specific path-type to layout this metadata. One path description for the data and another for the meta-information. However, because our path contains jumps, we can already achieve this functionality. Our solution is to append this information to the end of the path. The path jumps to each location to place text, in the same way as we describe a scatterplot visualisation, see \cref{fig:scatterPlot}. Alternatively labels could be added when the objects are placed -- where the object becomes a multivariate object (containing geometry and text). The position of the text label could be determined by one of the numerous label placement strategies~\cite{Christensen:1995:ESA:212332.212334, AGARWAL1998209}. Another way could be to couple the idea of logic within the path with label placement. Legends may also be placed along another path with decisions made at on construction or placement at any part of the model.

Our model allows the creation of many visualisations --  some good, some bad -- and by using the gene we are able to swap in/out parameters to discover or create new visualisations. Therefore, as one of the evaluation participants noticed, it is possible to create unsuitable visualisations. While not the focus of this paper, we do note that we are currently performing a more in depth evaluation of different visualisations designed by our method. With our builder interface, and even using the given (non- user-created) paths, we can combine and create thousands of parameters, and among them some will be poor visualisations. For instance, let's assume we want to apply a circle to a straight line path. By default, our circle segment has no data, and so it has a zero height; at the start of the process it will not necessarily look good. By adding default and random data we acknowledge that some early designs may not look good, and require effort on the part of the designer. But this is a typical solution to designing: ``\textit{the best way to a good idea is to have lots of ideas}'' (Pauling, \cite{Crick_on_LinusPauling}). Subsequently, one solution could be to include design rules into our path model. Rules could make some automatic design decisions to the placement of objects such as used in graph-drawing or label placement algorithms~\cite{AGARWAL1998209} or perhaps delivered from machine learning, such as Draco~\cite{MoritzDominik2019FVDK}.

In fact, this is one of the areas we want to expand. In future work we imagine automatically creating different genes, that will subsequently  produce different visualisations. We aim to further explore the effectiveness of visualisations in a more in depth evaluation and use these evaluations to drive future gene generation to recommend good visualisations to the user. An important output from these future evaluations would also be to establish recommended envelope parameters for given paths. This concept of a gene allows us to think of visualisations as a small series of properties. With this we would like to explore the possibility of gene mutation and combination. Borrowing concepts from Genetic Algorithms, we would like to create an implementation which allows users to combine and refine visualisations creating new generations and allowing for the usage of mutations. Another future work idea is to develop into 3D and  investigate different coordinate systems.

In this paper we have introduced a path-based design model, the Genii system (component library, builder and renderer).  We also presented an in-depth description of the flowmodel. The model encourages people to think about the visualisation in terms of a visual path, define a flowpath, apply different envelopes, place objects and filters on the path to generate alternative visualisation designs. The visualisations are saved as SVG, which can used in other applications; and have demonstrated how the system can be applied for use on a smartwatch. We demonstrate the model and system works, by several case study applications. Where we re-create several visualisations in the literature and one from a sketch. We have performed a user evaluation, which supports that the model is understandable and can be used to create a range of visualisations. By encoding the path model into a gene we are able to swap in/out parameters to quickly create different visual depictions. Finally, we hope that our path model will be extended and developed by other researchers. 
    
\section*{Supplemental Materials}
\label{sec:supplemental_materials}
\href{https://jamesjacko.github.io/genii/}{The Genii system (jamesjacko.github.io/genii/)} is released under CC BY 4.0, and includes the component library \href{http://doi.org/10.5281/zenodo.12571856}{doi:10.5281/zenodo.12571856},
renderer \href{http://doi.org/10.5281/zenodo.12571917}{doi:10.5281/zenodo.12571917} and builder \href{http://doi.org/10.5281/zenodo.12571944}{doi:10.5281/zenodo.12571944}. 

\section*{Figure Credits}
\label{sec:figure_credits}
Credit all images to authors; \cref{fig:teaser,FIG:Reilly,fig:flowgraph,fig:ModelDifferentPaths,fig:envelope,fig:Circular,fig:streamgraph,fig:envelopeOverlap,fig:scatterPlot,fig:differntCurves,fig:mobileTool,fig:caseStudy,FIG:GeniiPict,fig:susscorebox}, with the addition of accompanying thumbnails from Blascheck et al.\ \cite{BlascheckETAL2018} (top) and Brehmer et al.~\cite{Brehmer2018} (middle) in \cref{fig:caseStudy}. 

\acknowledgments{
This work is funded by KESS 2. Knowledge Economy Skills Scholarships (KESS 2) is a pan-Wales higher level skills initiative led by Bangor University on behalf of the HE sector in Wales. It is part funded by the Welsh Government’s European Social Fund (ESF) convergence programme for West Wales and the Valleys.%
}

\bibliographystyle{abbrv-doi-hyperref-narrow}

\bibliography{genii_REFS}           

\end{document}